\documentstyle[prb,aps,epsf]{revtex}

\newcommand{\be}{\begin{equation}}
\newcommand{\ee}{\end{equation}}
\newcommand{\bea}{\begin{eqnarray}}
\newcommand{\eea}{\end{eqnarray}}

\input epsf

\begin{document}
\draft
\title{Leaky interface phonons in AlGaAs/GaAs structures.}
\author{I. V. Ponomarev\cite{byline} and A. L. Efros}
\address{Department of Physics, University of Utah, Salt Lake City,\\
UT 84112}
\date{\today}
\maketitle

\begin{abstract}
A dispersion equation for the interface waves is  derived for the
interface of two cubic crystals in the plane perpendicular to [001]. A
reasonable hypothesis is  made about the total number of the acoustic
modes. Due to this hypothesis the number is 64, but not all of the modes
have physical meaning of the interface waves. The rules have been worked out
to select physical branches among all 64 roots of dispersion equation. The
physical meaning of leaky interface waves is discussed. The calculations
have been made for the interface Al$_{0.3}$Ga$_{0.7}$As/GaAs. In this case
all  physical interface modes have been shown to be leaky. The
velocities of the interface waves are calculated as a function of an angle
in the plane of interface. The results support a recent interpretation of a
new type oscillations of magnetoresistance as a resonant scattering of
two-dimensional electron gas by the leaky interface phonons.
\end{abstract}

\pacs{68.35.Ja, 62.65.+k, 73.50.Rb}

\section{Introduction}

\label{sec:1} The propagation of acoustic surface and interface waves has
attracted a significant  attention over the last decades. The concept of the
surface waves goes back to the famous paper by Lord Rayleigh\cite{rayleigh}.
The interface wave is a simple generalization of the surface wave, when the
second medium is not vacuum, and the wave propagates along the boundary
between two media. The theoretical study of the interface waves was
initiated by Stoneley\cite{ston1} who considered the case of two isotropic
solids.

 In anisotropic materials the interface waves  between hexagonal
crystals\cite{dj2,vel2} have been theoretically studied in sufficient
details. Relatively little known about effects of crystalline anisotropy
when the interface is formed by cubic crystals. To the best of our knowledge
the only numerical search for true interface wave velocities for several
combinations of the materials has been performed so far \cite{velasco,lim}.

Both surface and interface waves were initially studied in the context of
seismological waves propagating in the Earth's crust\cite
{sh2,scholte,phinney,pilant}. Later on these waves have been studied
experimentally in semiconductors by the light scattering\cite{sander,bril}.

The earlier theoretical studies by Lord Rayleigh and Stoneley ( see also\cite{LL}%
) prescribe to consider only those roots of the secular equations which give
an exponential decay of the surface wave in the medium under the surface and
an exponential decay of the interface wave in both media away from the
interface.

Probably, Phinney\cite{phinney} was the first to consider the so called
``leaky'' or ``pseudo'' waves which do not obey this prescription.
 Surface leaky
waves have been widely studied theoretically for both isotropic and
anisotropic crystalline materials (see brilliant review by Maradudin\cite
{Mar2} and references therein). To the best of our knowledge leaky interface
waves were studied only for isotropic case\cite{phinney,pilant} and
hexagonal crystals\cite{djafari}.

The interest to the leaky interface waves is stimulated by the fact that
true interface waves exist inside a very narrow range of the parameters.
Therefore in general case interface waves are leaky. This is not the case
for the surface waves  where the true non-leaky mode always exists in a
wide range of parameters\cite{limfar}. However, the dispersion equation for
surface waves also has several roots which give leaky solutions.

The structures with two-dimensional electron gas (2DEG), like
heterostructures or quantum wells provide another source of an interest to
the interface waves. The study of the interaction of 2DEG with surface waves
has been investigated long ago\cite{ing1,willet}. If 2DEG is far from the
surface, the electrons may interact with interface waves. Say, the electrons
may be scattered by thermally excited interface waves. This scattering
should not be less than scattering by bulk phonons, since in the vicinity of
the interface the three-dimensional densities  of the bulk and the interface
phonons are  of the same order. In the paper\cite{zud} we explained the novel
oscillations of magnetoresistance, observed in high-mobility 2DEG in
GaAs-AlGaAs heterostructures, by magneto-phonon resonance originating from
interaction of the 2DEG with thermally excited leaky interface acoustic
phonon modes.

The primarily goal of this paper is calculation of the interface waves for Al%
$_{0.3}$Ga$_{0.7}$As/GaAs interface on the basal (001) face. This is exactly
the interface used in the paper\cite{zud}. We have shown that all interface
waves in this case are leaky.

To this end we have derived analytically the secular equation for phase
velocity $v$ of the waves at the interface between two cubic crystals. We
have discussed the selection rules for the modes and have given a novel
general qualitative picture of the leaky interface waves. In this picture we
consider the conservation of energy and show that the amplitude of the wave
never becomes infinite if the problem is properly formulated. We show that
at some conditions leaky waves does not differ substantially from the true
waves. Finally we have obtained the numerical results, which were partially
used in Ref. \cite{zud}. 

The paper is organized as follows. The basis of the method is outlined in
section \ref{sec2}. In the third section we discuss general properties of
the secular equation, the selection rules for its solutions, and the
physical meaning of leaky waves. The numerical results and discussion are
presented in section \ref{num}. Finally, some auxiliary technical material
regarding calculations is given in the Appendices.

\section{General Formulation}

\label{sec2} Within the framework of the linear theory of elasticity the
equations of motion of the infinite medium are 
\begin{equation}  \label{eqmot}
\rho\frac{\partial u_i}{\partial t^2}= \frac{\partial\sigma_{ij}}{\partial
x_j}, \quad i=1,2,3,
\end{equation}
where $\rho$ is the mass density of the medium, $u_i({\bf r},t)$ is
Cartesian component of the displacement of the media at the point ${\bf r}$
at time $t$, and $\sigma_{ij}({\bf r},t)$ is the stress tensor. The latter
is given by Hooke's law 
\begin{equation}  \label{sigma}
\sigma_{ij}=\lambda_{ijkl}\frac{\partial u_{k}}{\partial x_l},
\end{equation}
where $\lambda_{ijkl}$ is symmetrical forth rank tensor. In cubic crystal
stress tensor can be conveniently written as 
\begin{equation}  \label{cubten}
\sigma_{ij}=C_{12}(\mbox{div}\ {\bf u})\delta_{ij}+C_{44}\left( \frac{%
\partial u_i}{\partial x_j}+\frac{\partial u_j}{\partial x_i}\right) +D\frac{%
\partial u_i}{\partial x_j}\delta_{ij},
\end{equation}
where summation over $i$ and $j$ is not assumed in the last term. Here $D$
is the anisotropy parameter: 
\begin{equation}
D=C_{11}-C_{12}-2C_{44}.
\end{equation}
For isotropic medium $D=0$.

In the following analysis we consider a system formed by two semi-infinite
cubic crystals. Elastic constants and density related to the lower part will
be denoted by prime symbol  (see Fig. 1). The interface is supposed to be on
(001) cut and perpendicular to $x_3$ axis. 

\begin{figure}[tbp]
\centerline{\epsfxsize=3.4in\epsfbox {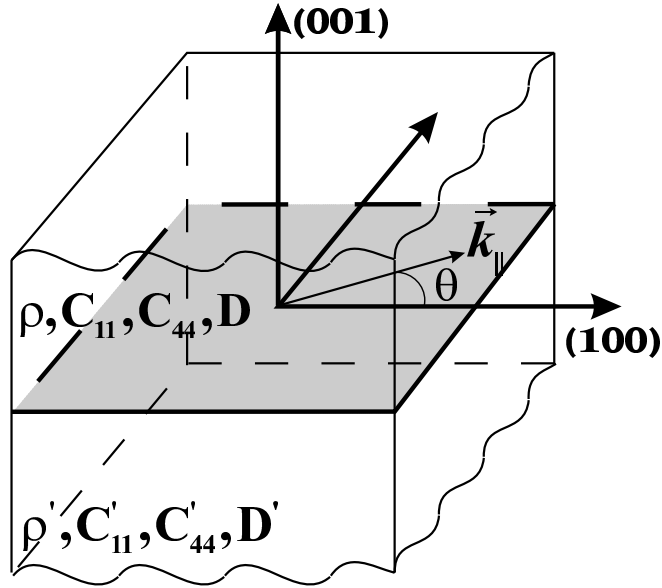}}
\caption{Structure for the study of interface acoustic waves.}
\label{fig1}
\end{figure}

The equations of motion (\ref{eqmot}) have to be supplemented by the
boundary conditions on the interface, expressing continuity of the
displacement and normal components of the stress tensor:
\begin{eqnarray}
u_i&=&u_i^{\prime}\left|_{x_3=0}\right.,\quad i=1,2,3,  \label{bc1} \\
\sigma_{i3}&=&\sigma_{i3}^{\prime}\left|_{x_3=0},\right.\quad i=1,2,3\ .
\label{bc2}
\end{eqnarray}

The homogeneous plane waves (bulk phonons) are the  simplest solutions
 of the wave equation in one infinite medium. They are 
\begin{equation}  \label{bulkph}
{\bf u}^{(l)}({\bf r},t)=\exp\left(i{\bf k}_{\|}{\bf x}_{\|}+ik_3x_3-
i\omega_{(l)} t\right)
\end{equation}
for three different branches $l=1,2,3$, where ${\bf x}_{\|}$ and ${\bf k}%
_{\|}$ are two-dimensional vectors with components $(x_1,x_2,0)$ and $%
(k_1,k_2,0)\equiv k(\cos\theta,\sin\theta,0)$ respectively (the angle $%
\theta $ is counted from [100] direction), and
 $\omega_{(l)}=s_{(l)}\sqrt{k^2+k_3^2}$ with the bulk sound velocity 
$s_{(l)}=s_{(l)}(\theta,\phi)$, where $\cos\phi=k_3/\sqrt{k^2+k_3^2}$.

The solution for the phase velocity can be determined by substitution the
plane wave of Eq. (\ref{bulkph}) into the equations of motion (\ref{eqmot}).
It gives the homogeneous set of linear equations. Setting the determinant of
the coefficients equal to zero, produces a cubic equation in $v^2$. Three
roots of this equation are squares of the velocities for three bulk phonons.

For the propagation along (001) plane one of the velocities $t_1= \sqrt{%
C_{44}/\rho}$ is independent of $\theta$ and represents a transverse mode.
Two others depend on angle of propagations in the plane. They are neither
longitudinal nor transverse, but we denote the upper branch by the letter ``$%
l$'' and the lower one by ``$t_2$''.

The interface between two half-infinite media introduces an inhomogeneity in 
$x_3$ direction. Therefore, we could expect that the plane waves {\bf also}
become inhomogeneous in this direction. The frequency and wave vector ${\bf k%
}_{\|}$ are the same in both media, but the component $k_3$ of wave vector
may be complex and different in upper and lower media. 
Moreover, since boundary conditions Eqs.(\ref{bc1},\ref{bc2}) comprise of 6
equations, the simplest solution for arbitrary direction of propagation
should consist of linear superposition of three terms described by Eq. (\ref
{bulkph}) for each medium with their own different complex components $k_3$.

The further analysis is facilitated by performing a rotation of the
coordinate frame in such a way that the direction of propagation of the
acoustic wave in the plane of interface is along the $x_1$ axis, {\em i.e.} $%
{\bf \tilde{k}}_{\|}=(k,0,0)$. Let 
\begin{equation}
\hat{\mbox{\bf T}}=\left[ 
\begin{array}{ccc}
\cos\theta & \sin\theta & 0 \\ 
-\sin\theta & \cos\theta & 0 \\ 
0 & 0 & 1
\end{array}
\right]
\end{equation}
be the transformation matrix which produces this rotation. Then, the
transformation law for the elements of the elastic modulus tensor under this
rotation is 
\begin{equation}  \label{tanlam}
\tilde{\lambda}_{ijkl}=\sum_{i^{\prime}j^{\prime}k^{\prime}l^{%
\prime}}T_{ii^{\prime}}T_{jj^{\prime}}T_{kk^{\prime}}T_{ll^{\prime}}
\lambda_{i^{\prime}j^{\prime}k^{\prime}l^{\prime}}.
\end{equation}


At the first stage of the analysis we determine the possible values of $k_3$
for given magnitude of $k$. To this end, we define $k_3\equiv ik\beta$, $%
\omega=kv$ and we suggest the solution for the interface waves in the
rotated system in the following form 
\begin{eqnarray}
u_i & =&A_ie^{-k\beta x_3}\exp\left[ik(x_1-vt)\right], \quad \mbox{for }
x_3>0,  \nonumber \\
u_i^{\prime}& =&A_i^{\prime}e^{k\beta^{\prime}x_3}\exp\left[ik(x_1-vt)\right]%
, \quad \mbox{for } x_3<0.  \label{anzats}
\end{eqnarray}
Conceptually, the $x_3$-dependence is the part of the ``amplitude'' (See
Ref. \cite{farnel}) and the wave-like properties are described by a common
propagation part $\exp\left[ik(x_1-vt)\right]$. Thus, the propagation vector
is always assumed to be parallel to the interface even though the exponent $%
\beta$ may be complex.

If Re$\beta,\beta^{\prime}>0$, then such a form describes a wave that
propagates in $x_1$ direction, whose amplitudes decays exponentially with
increasing distance into the medium from interface. The waves with (i) 
Re$\beta <0$ and Im$\beta <0$, (ii) Re$\beta^{\prime}<0$ and
 Im$\beta^{\prime}<0$, (iii) Re$\beta ,\beta^{\prime}<0$ and
 Im$\beta ,\beta^{\prime}<0$ are the leaky waves
which radiate the energy outward the interface.

 Substituting Eq.(\ref{tanlam}%
,\ref{anzats}) into the equations of motion Eq. (\ref{eqmot}) yields the set
of homogeneous equations for each media 
\begin{eqnarray}  \label{homset}
L_{ij}(v,\beta)A_j &=& 0.  \nonumber \\
L^{\prime}_{ij}(v,\beta^{\prime})^{\prime}A_j^{\prime}&=& 0,
\end{eqnarray}
where matrix ${\bf L}$ (or ${\bf L}^{\prime}$) has the form 
\begin{equation}  \label{matrix}
\left[ 
\begin{array}{ccc}
-C_{44}\beta^2+C_{11}-\frac{1}{2}D\sin^2 2\theta-\rho v^2 & -\frac{1}{4}%
D\sin 4\theta & \pm i\beta (C_{11}-C_{44}-D) \\ 
-\frac{1}{4}D\sin 4\theta & -C_{44}\beta^2+C_{44}+\frac{1}{2}D\sin^2
2\theta-\rho v^2 & 0 \\ 
\pm i\beta (C_{11}-C_{44}-D) & 0 & C_{44}-C_{11}\beta^2-\rho v^2
\end{array}
\right].
\end{equation}
The sign plus (minus) corresponds to the upper (lower) medium, and we
omitted primes for the lower medium. In each medium Eqs. (\ref{homset}) have
nontrivial solutions if the corresponding determinant of the coefficients
vanishes: 
\begin{equation}
{\rm det}({\bf L}) =0,  \label{detL}
\end{equation}
It gives the secular equation on unknown values of $\beta$ with a phase
velocity $v$ as a parameter. The explicit form of Eq. (\ref{detL}) is given
in Appendix \ref{ApC}. Due to the fact that we are seeking the solution for
wave propagation in crystal plane of mirror symmetry this equation is
bicubic in $\beta$ and the roots have inversion symmetry with respect to the
origin of the complex plane\cite{farnel}. One can also show that if $\beta_l=%
{\beta_R}_j+i{\beta_I}_j$ at $j=1,\ldots ,6$ are the roots of Eq. (\ref{detL}%
) with complex velocity $v=v_R+iv_I$, then the roots $\beta_j={\beta_R}_j-i{%
\beta_I}_j$ are the roots of the same equation with $v=v_R-iv_I $. Here the
superscripts ``$R$'' and ``$I$'' denote the real and imaginary parts
respectively. %

The amplitudes $A_{\alpha}$ (or $A_{\alpha}^{\prime}$) for any $\beta_j$ ($%
\beta_j^{\prime}$) are related by 
\begin{equation}  \label{ampcon}
\frac{A_1^{(j)}}{C_1^{(j)}}=\frac{A_2^{(j)}}{C_2^{(j)}}= \frac{A_3^{(j)}}{%
C_3^{(j)}}=K_j, \qquad j=1,\ldots,6,
\end{equation}
where the ${K_j}$ are constants and $C_{\alpha}^{(j)}(v,\beta_j)\
(\alpha=1,2,3)$ are the cofactors of the elements in the first row of the
matrix {\bf L}: 
\begin{eqnarray}
C_1^{(j)}&=&L_{22}L_{33},  \nonumber \\
C_2^{(j)}&=&-L_{21}L_{33},  \nonumber \\
C_3^{(j)}&=&-L_{31}L_{22}.  \nonumber
\end{eqnarray}

The next step of our analysis is a construction of the general solution,
which satisfies boundary conditions Eq. (\ref{bc1},\ref{bc2}). To this end,
we form a linear combinations from three terms (\ref{anzats}) with undefined
constants $K_j$ and $K^{\prime}_j$ for each medium: 
\begin{eqnarray}
u_{\alpha} & =&\sum_{j=1}^3\frac{C_{\alpha}^{(j)}}{C_{1}^{(j)}}K_j
\exp\left(ik(x_1-vt+i\beta_j(v) x_3)\right),\quad \mbox{for } x_3>0, 
\nonumber \\
u_{\alpha}^{\prime}& =&\sum_{j=1}^3\frac{C_{\alpha}^{\prime(j)}}{C_{
1}^{\prime(j)}} K_j^{\prime}\exp\left(ik(x_1-vt-i\beta^{\prime}_j(v)
x_3)\right) ,\quad \mbox{for } x_3>0.  \label{fulsol}
\end{eqnarray}
Substitution of this form for the displacement field into Eqs. (\ref{bc1}, 
\ref{bc2}) leads to a set of 6 (in general case) homogeneous linear
equations for the ${K_j,K_j^{\prime}}$. The nontrivial solutions exist if
the corresponding determinant vanishes: 
\begin{equation}  \label{drg}
|D_{kl}^{(\gamma)}(v)|=0,\quad k,l=1,\ldots ,6.
\end{equation}
Eq. (\ref{drg}) is the dispersion relation for the phase velocity $v$ of the
interface acoustic wave. In general it has to be solved numerically. The
left hand side function $D(v)\equiv |D_{kl}^{(\gamma)}|$ is some algebraic
expression. Therefore, in general, the roots of Eq. (\ref{drg}) are complex.
Moreover, since $D(v)$ comprises of six different decay constants $%
\beta_j(v),\beta^{\prime}_j$, which involve square roots from some
expressions of $v^2$, the function $D(v)$ is multi-valued analytical
function of complex variable $v$, defined on its associated Riemann sheets.
The upper-script $\gamma$ enumerates these sheets. The number of Riemann
sheets are determined by different combinations of $\beta$ branches.
However, not all from $6!/3!3!=20$ combinations of $\beta_j$ are possible 
 for each medium in the superpositions of Eq. (\ref{fulsol}). Each of
three decay constants must be taken from the different roots $\beta^2$ of
cubic equation ${\rm det}({\bf L})=0$ at fixed $v^2$. Therefore, the total
number of possible combinations is $2^3\times2^3=64$. This is the number of
Riemann sheets for our case. Since simultaneous change of all signs $%
\beta,\beta^{\prime}$ in Eq. (\ref{drg}) does not change the form of
determinant (see Appendix A), it is enough, in fact, to investigate 32
independent Riemann sheets in order to find all possible roots of the
dispersion relation. In the isotropic case and for the propagation along the
directions of high symmetry the number of independent Riemann sheets reduces
to 8.

The sign convention for the sheets is determined by real part of $\beta$. It
is denoted as follows: 
\begin{equation}  \label{scon1}
\left( \mbox{sign}\mbox{Re}(\beta_1),\mbox{sign}\mbox{Re}(\beta_2), %
\mbox{sign}\mbox{Re}(\beta_3), \mbox{sign}\mbox{Re}(\beta^{\prime}_1),%
\mbox{sign}\mbox{Re}(\beta^{\prime}_2), \mbox{sign}\mbox{Re}%
(\beta^{\prime}_3)\right).
\end{equation}
Let us assume that $\gamma=1$ corresponds to the case (++++++). If a
solution exists on this sheet, then it is a true interface wave, which is
also called Stoneley wave. All other $\gamma$ correspond either leaky waves
or non-physical solutions. Some of them may also correspond to bulk phonons
(see section \ref{sec3}). In the next section we formulate selection rules
for physical solutions. The right direction of the energy flux is the main
principle for the selection. The explicit form of Eq. (\ref{drg}) and the
way of enumeration of its Riemann sheets are given in Appendix \ref{ApC}.


\section{Selection rules for velocity and physical meaning of leaky waves}

\label{sec3}

The question of total number of possible values of velocity has been
investigated in earlier 70s theoretically\cite{ansell} for the isotropic
solid -- liquid interface, and numerically for the case of two isotropic
solids\cite{pilant}. In the case of the liquid -- solid interface there are 
{\em eight} Riemann sheets. It is shown\cite{ansell} that
the roots on all these sheets are the roots of an eight order
polynomial in $v^2$ with real coefficients, and so there are {\em eight}
complex roots which are either real or come  in complex conjugate pairs.
Numerical investigation of the Stoneley equation (\ref{stdr1}) for isotropic
solid -- solid interface\cite{pilant} has shown that there are {\em sixteen}
independent roots on its {\em sixteen} Riemann sheets.

Thus, we can put forward a simple hypothesis: The number of possible values
of $v^2$ is equal to the number of Riemann sheets. However, some of them may
be degenerate so we are speaking about {\em maximum} number of different $%
v^2 $. Note that this hypothesis is true for isotropic surface wave either.
It follows from it that the maximum number of modes in our interface is 64.
To check this hypothesis we have calculated this number for one of direction
which does not have any special symmetry. The result is 64.

Let us turn now to the problem of roots classification for the case $%
\gamma>1 $. First of all we discuss real roots for $v$ of the dispersion
equation Eq. (\ref{drg}). If all $\beta$ are pure imaginary, it corresponds
to refraction of bulk phonons and has nothing in common with leaky waves. If
at least one of $\beta$ at any side has a negative real part, such solution
should be considered as non-physical. If it happens that some $\beta$ are
imaginary but some are complex with positive real part, then it relates to
the problem of total internal reflection of bulk phonons. In numerical
analysis we discard such solution, since they have a different nature.

Now we come to complex $v$. Let us assume that we have real positive
frequency $\omega>0$ and complex root 
\begin{equation}  \label{cvel}
v=v_R\mp iv_I
\end{equation}
of the Eq. (\ref{drg}). As follows from Appendix A all complex roots form
such pairs. Then, the wave vectors of propagation along $x_1$-axis for these
solutions will be complex and equal to 
\begin{equation}
k=\omega/v=\omega/(v_R\mp iv_I)=k_R\pm ik_I.
\end{equation}
If $k_R>0$, it corresponds to the running wave propagating from the left to
the right with exponentially decreasing or increasing amplitudes. Since both
of these solutions always meet in pairs we will consider only the wave
attenuated from the left to the right. Then we should chose only the root $%
k_R+ ik_I$ or $v_R-iv_I$, where $v_R,v_I,k_R,k_I>0$. 

Since $\gamma>1$, one or several $\beta$ have negative real part, i. e. 
\begin{equation}  \label{betadif}
\beta=-\beta_R-i\beta_I,
\end{equation}
where $\beta_R>0$ and sign $\beta_I$ is not determined yet.

It is useful to introduce the following notations: 
\begin{eqnarray}
\tilde{v}_R &=&(v_R^2+v_I^2)/v_R, \\
\tilde{\beta}_R &=&\beta_R-\beta_Iv_I/v_R, \\
\tilde{\beta}_I &=&\beta_I+\beta_Rv_I/v_R,
\end{eqnarray}
After substitution $k=\omega/v$ the term has a form of inhomogeneous plane
wave (see also the succeeded section) 
\begin{equation}  \label{ipw}
e^{-i\omega (t-x/\tilde{v}_R-z\tilde{\beta}_I/ \tilde{v}_R)}e^{-\omega/%
\tilde{v}_R(xv_I/v_R-z\tilde{\beta}_R)}.
\end{equation}
In fact, the sign of $\beta_I$ is not arbitrary, but is dictated by the
radiation condition\cite{Ingeb,Mar1}. Indeed, since we consider here
lossless media the only reason for the amplitude attenuation along direction
of propagation on the interface is 
radiation of energy away from the interface
into the bulk media. The total wave vector ${\bf q}=\omega/\tilde{v}_R(1,0,%
\tilde{\beta}_I)$ is no longer parallel to the boundary, but is inclined to
it, which indicates the presence of continuous flow of energy from the
boundary to the bulk. Note that direction of phase propagation ${\bf q}/q$
does not represent the direction of energy flow itself. The later is
determined by the time averaged power flux 
\begin{equation}  \label{flux}
W_{\alpha}=-$Re$\langle\frac{1}{2}\sigma_{\beta\alpha}\dot{u}%
_{\beta}^*\rangle, \qquad \alpha,\beta=1,2,3.
\end{equation}
One can show that for small $v_I$ the sign of the component $W_3$ which
determines the flow perpendicular to the interface coincides with the sign
of $\tilde{\beta}_I$. Therefore, in notations of Eq. (\ref{betadif}) the
sign $\tilde{\beta}_I$ must be positive to guarantee the proper radiation
condition.


Now we come to the physical meaning of the leaky waves. Discussing Rayleigh
waves Landau and Lifshitz\cite{LL} prescribe to drop as non-physical the
solution, which increases away from the surface. In the theory of leaky
waves we take into consideration such a solution. The goal of this part is
to give physical explanation of leaky waves (see also review by Maradudin 
\cite{Kress}).

On both side of the interface our solution consists of three inhomogeneous
plane waves. At large values of $x$ and $z$ these waves can be considered as
independent. So we concentrate on one of them choosing the wave at $z>0$
with negative real part of $\beta$, i.e. the wave exponentially increasing
with $z$. This wave has a form Eq. (\ref{ipw}). 
\begin{equation}  \label{int1}
e^{-i\omega (t-x/\tilde{v}_R-z\tilde{\beta}_I/ \tilde{v}_R})e^{-\omega/%
\tilde{v}_R(xv_I/v_R-z\tilde{\beta}_R)},
\end{equation}
where $\tilde{\beta}_R>0$ and $\tilde{\beta}_I>0$. This is an inhomogeneous
bulk plane wave with wave vector ${\bf q}=\omega/\tilde{v}_R(1,0,\tilde{\beta%
}_I)$ propagating away from the interface. The lines of constant phase are
determined by the equation $t-x/\tilde{c}_R-z\tilde{\beta}_I/\tilde{c}_R=C_1$%
, and the lines of constant amplitude are given by the equation $xv_I/v_R-z%
\tilde{\beta}_R=C_2$ (See Fig. 2) The later expression shows that such modes
attenuate when they move along the surface $z=0$. 

\begin{figure}[t]
 \vspace{1cm}

  \centerline{\hbox{
    \epsfxsize=3.40in
      \epsfbox [54 365 560 685] {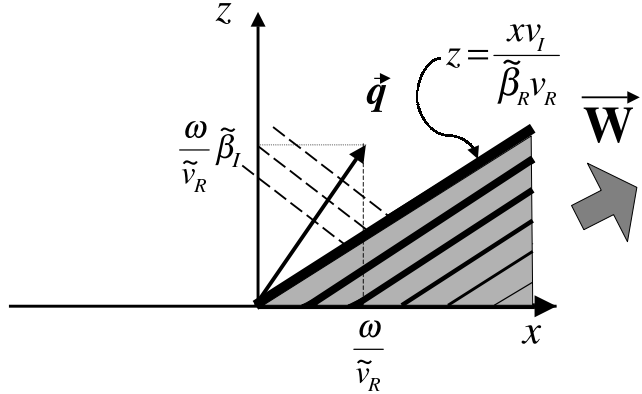}
                    }
        }

\caption{Illustration for geometry of leaky interface wave term. Power flux $%
{\bf W}$ and wave vector ${\bf q}$ are shown. The full lines with different
thickness are the lines of a constant amplitude. The thickness of the lines
indicates schematically the absolute value of the amplitudes. Dashed lines
are the lines of a constant phase. For visual clarity the angle of constant
amplitude lines has been exaggerated.}
\label{fig2}
\end{figure}

The central point of our understanding of the leaky waves is that one can
not consider this waves in whole region of $x$, since amplitude becomes
infinite when $x\longrightarrow -\infty$. This happens 
 because we have chosen the
solution which propagates from the left to the right along $x$-axis. Thus,
we should propose that the wave creates at some  line, say $x=0$, in the plane
of the interface. Our equations of motion do not include any dissipation,
therefore the attenuation in the plane of interface may be only due to
radiation into the media. There is an important theorem \cite{potok} stating
that for inhomogeneous plane waves the energy flux is parallel to the plane
of constant amplitude. The cross-sections of these planes with the plane $zx$
are shown by full lines of different thickness. The upper line is thicker
because the amplitude of the wave at the line $x=0$ is the largest and it
decrease with increasing $x$. That is why the amplitude at any point $x$
increases with $z$ at $z<xv_I/\tilde{\beta}v_R$. It follows from the above
theorem that the energy flux can not cross the planes of constant amplitude.
It also can not cross the plane of maximal amplitude. This means that all
the wave is within the wedge formed by the plane of interface and the plane
of maximal amplitude $z=xv_I/\tilde{\beta_R}v_R$, at least in a sense that
the whole energy of the wave is within this wedge. The amplitude of the wave
is finite everywhere in this region.

One gets a severe contradiction considering a stationary problem. Indeed, 
the equation (\ref{int1}) gives nonzero result outside the wedge as well.
Moreover amplitude diverges when $z$ tends to infinity. This is an artifact
of the stationary consideration. The origin of the  divergence is the 
 infinite amplitude of the
wave at point $x=-\infty$. The increase of the amplitude at large $z$ is an
artifact originated from the flux coming from large negative $x$.

It is important to mention that the leaky wave does not differ substantially
from the true interface wave only if $k_R\gg k_I$ in Eq. (18) or $v_R\gg v_I$%
. Since $\beta_R$ is not small, it means that the angle between the planes,
forming the wage in Fig. 2, should be small. If this condition is not
satisfied, the interface (or surface ) wave can not be considered as a wave
since the wavelength is larger than the attenuation length.

The problem does not contain any small parameter which could make this
condition fulfilled.
The numerics show, however, that the majority of modes have small
attenuation. The roots  of  this phenomenon are  not clear for us.

Thus, based on discussion in this section we use the following selection
rules for values $v=v_R-iv_I$, which are solutions of Eq. (\ref{drg}).

\begin{enumerate}
\item  $v_R>0$, $v_I\ge 0$.

\item  If $v_I=0$, than Re$\beta$, Re$\beta^{\prime}>0$.

\item  If $v_I>0$ and Re$\beta<0$, than Im$\beta<0$.

\item  If $v_I>0$ and Re$\beta^{\prime}<0$, than Im$\beta^{\prime}<0$.
\end{enumerate}

\section{Numerical results}

\label{num}

To calculate velocities as a function of angle we use Eq. (\ref{efdetL}) for 
$\beta^2$ and Eq. (\ref{detvan}). At the first step we divide complex plane $%
v$ in interval 2 km/s$<v_R<7$ km/s, $0<v_I<0.6$ km/s into 400 squares. For
the vertex of each square we find six values of $\beta$ for upper medium and
six values of $\beta^{\prime}$ for the lower medium using Eq. (\ref{efdetL}%
). For each value of $v$ we find 32 different combinations of $\beta$ (each
of 6 values) and substitute them into Eq. (\ref{detvan}). For each Riemann
sheet $\gamma$ we find all minima of abs($|D_{ij}^{(\gamma)}(v)|$) with
respect to $v_R$ and $v_I$ using a standard program. For those minima which
are close to zero, we do iterative search of roots. The parameters of the
bulk lattices have been taken from the Table I. 
\parbox{18cm}{
\begin{table}\label{tbl1}
\caption{Densities(g/cm$^3$), elastic constants ($10^{10}$N/m$^2$), and sound 
velocities (km/s) in the directions of high symmetry for bulk 
crystals\protect{\cite{simon}}.}
\begin{tabular}{|ccccccccc|}
Crystal & $\rho$ & $C_{11}$& $C_{12}$& $C_{44}$ &$l_{[100]}$& $t_1$ &
$l_{[110]}$ & $t_{2[110]}$ \\
\hline
Al$_{0.3}$Ga$_{0.7}$As&4.794&12.24&5.65&5.90& 5.05&3.51&5.56&2.62 \\
GaAs                  &5.307&12.26&5.71&6.0 & 4.81&3.36&5.31&2.48 \\
\end{tabular}
\end{table}
}
  To check the method some results have been obtained using completely
different Surface Green Function Matching method\cite{GM,vel2}, which is
discussed in Appendix B. We have not found any differences between the
results of two methods.

For an additional check of the problem we have calculated true surface waves
for both materials. The dispersion relation for surface waves on (001) cut
of cubic crystals can be obtained from the determinant of truncated
interface matrix Eq. (\ref{Dmatrix}). For the upper (lower) medium we should
take lower left (right) 3x3 part of the matrix. We have gotten 2.873 km/s
(2.737 km/s) for GaAs (Al$_{0.3}$Ga$_{0.7}$As). These results may be
compared with the results by Farnell\cite{farnel} Our results are slightly
different because of the difference in the parameters of bulk materials.
Taking parameters used by Farnell we have obtained his velocities with a
very high accuracy.

\begin{figure}
  \centerline{\hbox{ 
    \epsfbox {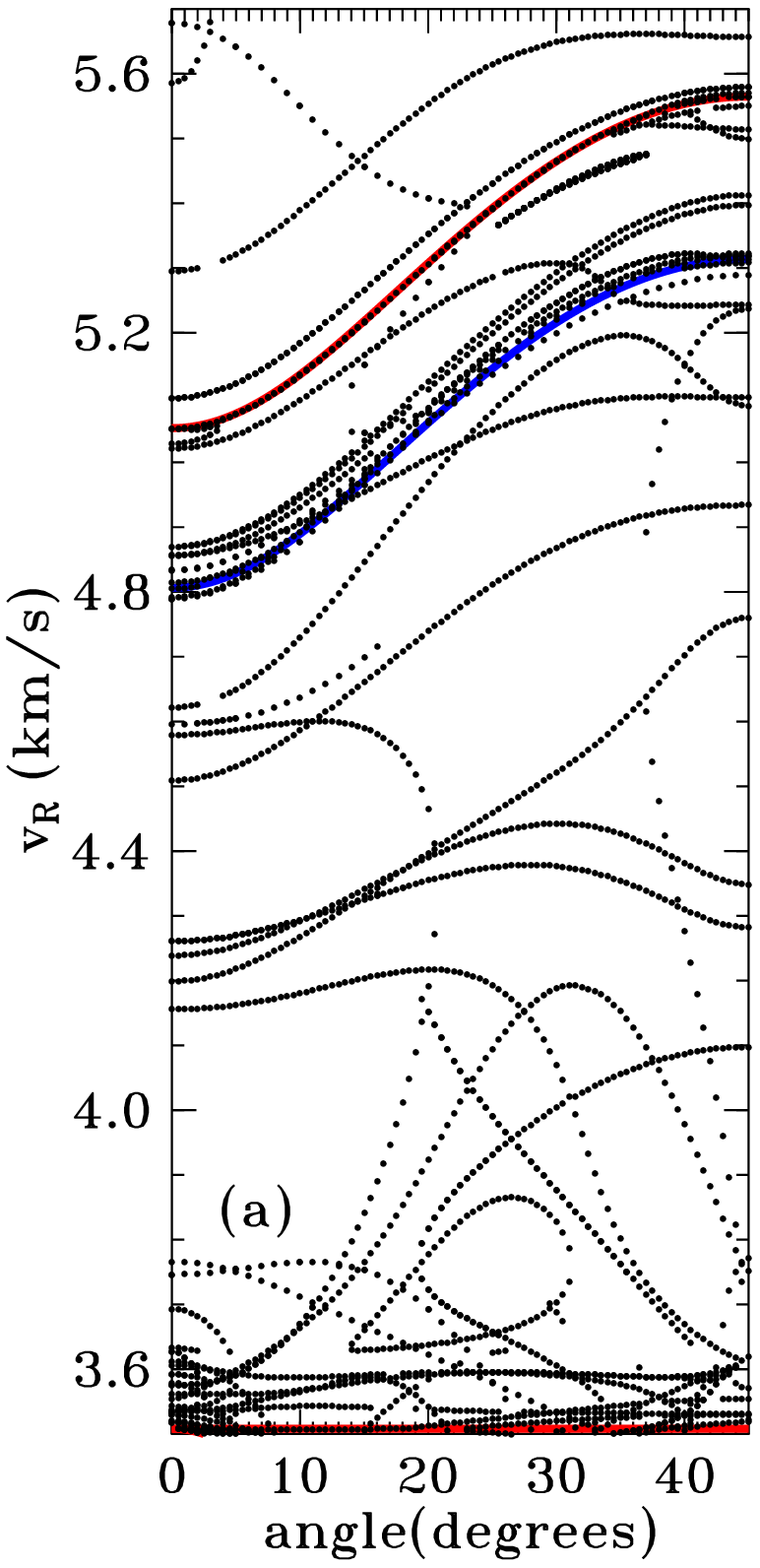}
    \hspace{-0.1in}
    \epsfbox {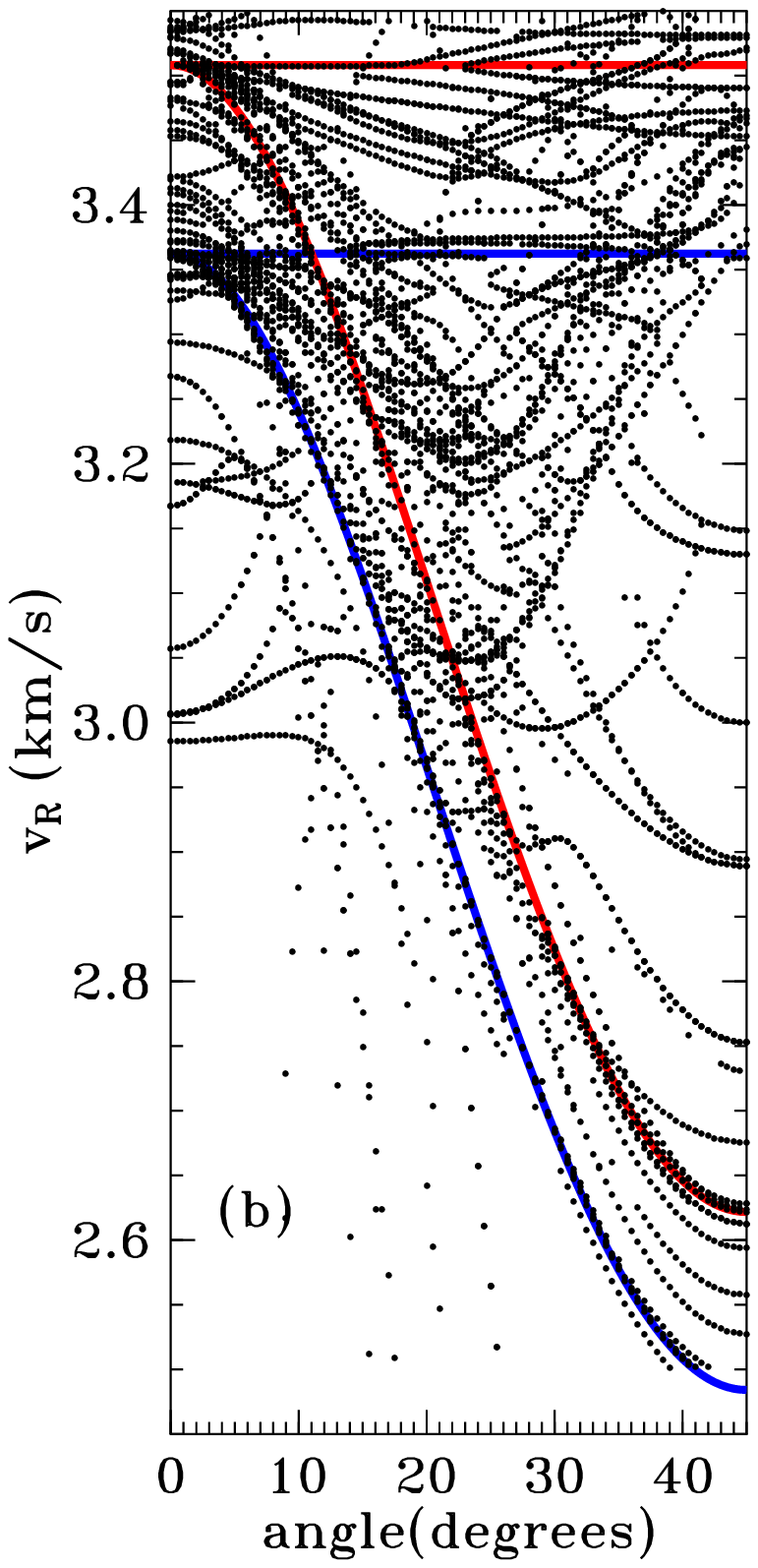}
    }
  }
\caption{The real parts of complex velocities for all leaky interface 
waves in the range 3.5 km/s $<v_{R}<$ 5.7 km/s {\bf (a)} and 
2.4 km/s $<v_{R}<$ 3.5 km/s {\bf (b)}. The solid lines present velocities of 
bulk acoustic waves for both media.
}
\label{fig3}       
\end{figure}

\begin{figure}[h]
\centerline{\epsfbox{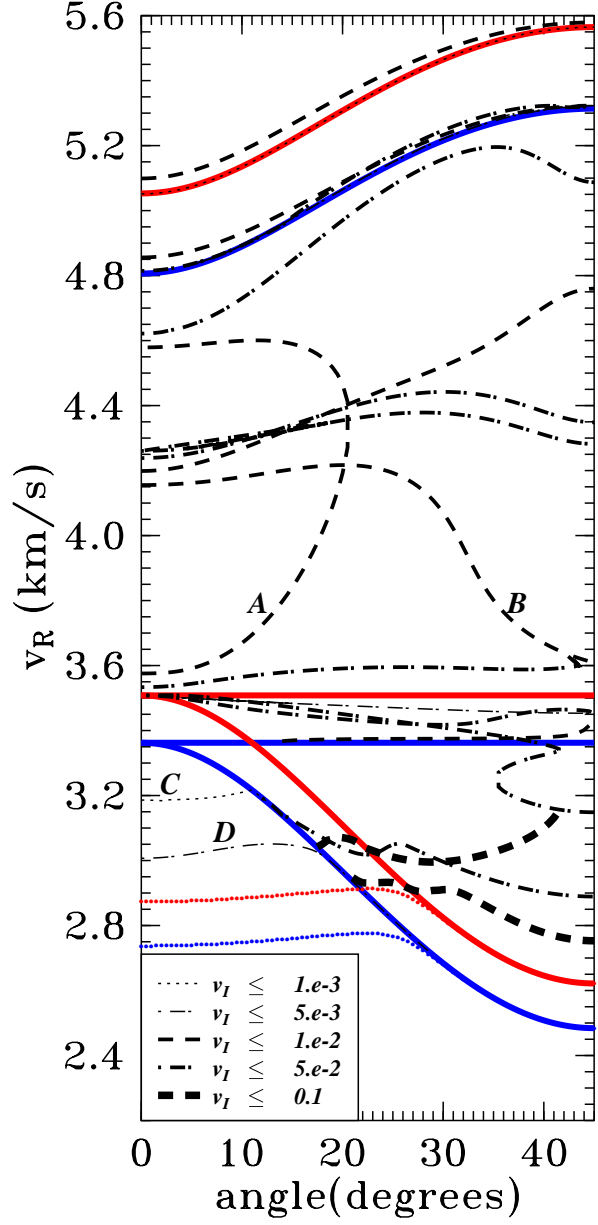}}
\caption{The real parts of complex velocities for the selected modes in the 
range 2.4 km/s $<v_{R}<$ 5.7 km/s. See explanation in the text. 
The numerical parameters for these modes are given in the Table II.
The value of attenuation is determined by the imaginary part of
complex velocity. Different linestyles and thickness correspond to 
different magnitude of average imaginary part $\overline{v_{I}}$ (km/s) of 
velocity. The solid lines present velocities of bulk acoustic waves and 
the dotted  lines are true surface acoustic waves for both media (Upper 
curves correspond to Al$_{0.3}$Ga$_{0.7}$As and lower curves correspond 
to GaAls).
}
\label{fig4}
\end{figure}

The results for leaky interface waves are shown in Fig. \ref{fig3}
and Fig. \ref{fig4}. Plotting
this figure we have taken into account all selection rules formulated at the
end of section \ref{sec3}. The discontinuities appear because at some points
these rules are not fulfilled and the corresponding modes become
non-physical. All the modes have different anisotropy, different attenuation
and different angle intervals of their existence.
In Fig.\ref{fig3}  we draw all the modes
(Note that velocity scales in 3(a) and 3(b) are  different). One can see that
there are abundant number of leaky interface modes in the velocity range 
between 3km/s and 4km/s. Thorough analysis shows however that a majority of 
those modes exists in a very restricted range of the angles. Probably, 
it will be difficult to detect them experimentally.
In Figure \ref{fig4}  we draw selected modes which exist in the
large interval of the angles only. 
We left one mode ({\it A}) with a small angle range  
in the figure as a typical example of discarded modes. 
For further clarification of  the picture we also discard all physical 
modes with a strong anisotropy of their real and/or imaginary parts of 
the velocity. One of such modes with strong real part anisotropy (B) is left 
 in the figure as an example. 

All numerical parameters for the modes presented in 
Fig. \ref{fig4} are given in the table II.
For modes classification we introduce the following parameters.
\begin{enumerate}
\item Averaged over angles real $\overline{ v_R}$ and imaginary
$\overline{v_I}$ part of the velocity.
\item Angle range parameter 
$\delta_{an}= (\varphi _{\max }-\varphi _{\min })\,4/\pi$. It equals
to unity when a mode exists on whole range of angles and less than one
otherwise.
\item Anisotropy parameters for real and imaginary parts of the velocity
\begin{equation}
\sigma _{v_{R}}=(v_{R\max }-v_{R\min })/\overline{v_R}, \qquad
\sigma _{v_{I}}=(v_{I\max }-v_{I\min })/\overline{v_I}.
\end{equation}
Modes with smaller anisotropy have smaller $\sigma$.
\end{enumerate}
Among the modes  there are two ({\it C} and {\it D} in the figure)
which remind those in the surface acoustic problem.
Namely,
for Al$_{x}$Ga$_{1-x}$As materials there are always\cite{farnel} true SAWs
(they are shown by dotted lines in the figure), 
which change very a little with
angle until  they meet the lowest bulk transverse velocity curves 
(they are shown by solid line). After that point the true SAWs repeat 
behavior of their bulk velocity curves up to the end - 45 degrees angle 
on the picture. 
Meanwhile, in the region between two transverse bulk velocity curves the 
leaky surface acoustic waves appear  with approximately the same real part
of velocity.

The situation for interface waves is  different. As we mentioned,
 there are no  true
interface waves. However, there are leaky interface waves with a very 
small attenuation at small angles  before ``colliding'' with the bulk 
transverse velocity curve. At larger angles the leaky modes
 acquire larger imaginary part of the velocity 
and become stronger attenuated (See Table II).

%
%
%
%
%

\section{\protect\bigskip Conclusion}

We have derived a dispersion equation for interface waves at the interface
of two cubic crystals in the plane, perpendicular to [001]. Analyzing
different solutions for the interface waves we come to conclusion that the
total amount of interface modes in each direction is equal to the number of
the Riemann sheets. In our case this is 64. We have successfully checked
this hypothesis by calculating the number of modes in one direction of the
interface plane, which does not have any special symmetry.

The computations have been made for the interface Al$_{0.3}$Ga$_{0.7}$%
As/GaAs. We have shown that in this case all interface modes, which have
physical meaning are leaky, but majority of them have small attenuation in
the direction of propagation. We show that for understanding of the physical
meaning of leaky waves one should consider not a stationary problem , but
the problem starting with creation of the wave at some line in the interface
plane.

After that we are able to formulate how to separate this 64 modes into
physical and non-physical modes. This separation mainly based upon some
theorems on the energy flux and upon an assumption that if a mode deviates
from the interface in some medium, the energy flux should go in the same
medium.

Using the elastic moduli of the bulk lattices we have performed numerical
calculations of the velocities of the interface waves as a function of an
angle in the plane of interface in a wide range of velocities. The results
are shown in Fig. \ref{fig3}. One can see two close groups of modes within
the intervals 3$-$3.5 km/s and 4.2$-$4.5 km/s respectively. These groups may
be responsible for two periods of oscillations which have been observed in
the experiment with the two-dimensional electron gas in magnetic field\cite
{zud}, mentioned in the introduction. 
Note that the velocities of the leaky interface waves may be sensitive to
the difference of the bulk media parameters. This difference is not known
good enough. This fact may be responsible for possible deviation of our
calculations from the experimental data.

\section{Acknowledgments}

The authors wish to thank B. Bromley for the helpful advises. IVP is
grateful to V. Velasco for giving his numerical source code for calculations
of Stoneley waves which was valuable check for our method. This work was
supported by Seed grant of the University of Utah. 

\parbox{18cm}{
\begin{table}[t]
\label{tabv}
\caption{Parameters for the selected modes.
The units for average velocities are km/s.
}
\begin{tabular}{|c|c|c|c|c|c|c|c|c|c|}
$\#$ & $\delta_{an}$ & $\overline{v_R}$ & $\overline{v_I}$ & $\sigma
_{v_{R}}$ & $\sigma _{v_{I}}$ & $v_{R\min }$ & $v_{R\max }$ & $v_{I\min }$
& $v_{I\max}$ \\ \hline
 $A$ & 0.46 & 4.166 & 6.5$\times 10^{-3}$ & 0.25 & 0.39 & 3.58 & 4.60 & 
5.2$\times 10^{-3}$ & 7.7$\times 10^{-3}$ \\ \hline
$B$ & 0.99 & 4.022 & 9.5$\times 10^{-3}$ & 0.15 & 2.48 & 3.60 & 4.22 & 
2.1$\times 10^{-4}$ & 2.4$\times 10^{-2}$ \\ \hline
 $C_{1}$ & 0.26 & 3.195 & 1.9$\times 10^{-4}$ & 0.01 & 2.65 & 3.18 & 3.22 & 
1.8$\times 10^{-11}$ & 5.1$\times 10^{-4}$ \\ \hline
$C_{2}$ & 0.73 & 3.000 & 3.3$\times 10^{-2}$ & 0.10 & 2.50 & 2.89 & 3.19 & 
3.8$\times 10^{-7}$ & 8.4$\times 10^{-2}$ \\ \hline
$D_{1}$ & 0.63 & 2.966 & 5.3$\times 10^{-3}$ & 0.11 & 1.00 & 2.72 & 3.05 & 
1.9$\times 10^{-3}$ & 7.2$\times 10^{-3}$ \\ \hline
$D_{2}$ & 0.53 & 2.856 & 6.6$\times 10^{-2}$ & 0.07 & 2.76 & 2.75 & 2.95 & 
2.3$\times 10^{-4}$ & 0.18 \\ \hline
1 & 1 & 5.369 & 7.9$\times 10^{-3}$ & 0.09 & 1.32 & 5.10 & 5.58 & 
3.4$\times 10^{-3}$ & 1.4$\times 10^{-2}$\\ \hline
2 & 1 & 5.167 & 4.2$\times 10^{-3}$ & 0.1 & 0.31 & 4.83 & 5.32 & 
3.7$\times 10^{-3}$ & 5.0$\times 10^{-3}$\\ \hline
3 & 1 & 5.330 & 2.1$\times 10^{-4}$ & 0.1 & 3.44 & 5.05 & 5.56 & 
2.2$\times 10^{-6}$ & 7.1$\times 10^{-4}$ \\ \hline
4 & 1 & 5.112 & 6.8$\times 10^{-3}$ & 0.09 & 0.79 & 4.86 & 5.32 & 
4.1$\times 10^{-3}$ & 9.4$\times 10^{-3}$\\ \hline
5 & 1 & 5.115 & 4.0$\times 10^{-3}$ & 0.1 & 0.19 & 4.83 & 5.32 & 
3.7$\times 10^{-3}$ & 4.5$\times 10^{-3}$ \\ \hline
6 & 1 & 5.104 & 3.5$\times 10^{-2}$ & 0.1 & 1.41 & 4.81 & 5.32 & 
9.5$\times 10^{-3}$ & 5.9$\times 10^{-2}$ \\ \hline
7 & 1 & 4.960 & 3.5$\times 10^{-2}$ & 0.12 & 1.72 & 4.62 & 5.20 & 
1.6$\times 10^{-2}$ & 7.7$\times 10^{-2}$\\ \hline
8 & 1 & 4.446 & 8.7$\times 10^{-3}$ & 0.13 & 1.75 & 4.20 & 4.76 & 
2.6$\times 10^{-3}$ & 1.8$\times 10^{-2}$\\ \hline
9 & 1 & 4.360 & 2.1$\times 10^{-2}$ & 0.05 & 0.65 & 4.24 & 4.44 & 
1.3$\times 10^{-2}$ & 2.7$\times 10^{-2}$\\ \hline
10 & 1 & 4.323 & 2.2$\times 10^{-2}$ & 0.03 & 0.58 & 4.26 & 4.38 & 
1.4$\times 10^{-2}$ & 2.7$\times 10^{-2}$\\ \hline
11 & 0.94 & 3.578 & 2.0$\times 10^{-2}$ & 0.02 & 2.22 & 3.53 & 3.60 & 
1.2$\times 10^{-3}$ & 4.6$\times 10^{-2}$ \\ \hline
12 & 0.99 & 3.475 & 3.1$\times 10^{-3}$ & 0.02 & 1.27 & 3.45 & 3.51 & 
2.0$\times 10^{-4}$ &4.1$\times 10^{-3}$ \\ \hline
13 & 0.99 & 3.452 & 3.1$\times 10^{-2}$ & 0.03 & 1.63 & 3.42 & 3.51 & 
2.0$\times 10^{-4}$ &5.1$\times 10^{-2}$ \\ \hline
14 & 0.68 & 3.77 & 8.8$\times 10^{-3}$ & 0.01 & 1.35 & 3.37 & 3.42 & 
8.6$\times 10^{-4}$ & 1.3$\times 10^{-2}$\\ \hline
15 & 0.99 & 3.79 & 4.1$\times 10^{-2}$ & 0.11 & 1.46 & 3.15 & 3.51 & 
2.0$\times 10^{-4}$ & 5.9$\times 10^{-2}$\\ \hline
16 & 0.53 & 3.040 & 6.9$\times 10^{-2}$ & 0.05 & 1.69 & 3.00 & 3.15 & 
1.0$\times 10^{-3}$ & 0.12\\ 
\end{tabular}
\end{table}
}
\appendix
\section{Explicit form of the Dispersion relation in general,
symmetry directions, and isotropic cases}\label{ApC}
From the  determinant (\ref{detL}) we obtain the following
equation on unknown variable $\beta$ with a phase velocity $v$ as a
parameter: 
\begin{equation}  \label{efdetL}
\beta^6-\beta^4\left[a+b+c-\frac{(\lambda^2-1-d)^2}{\lambda^2}\right]+
\beta^2\left[ab+bc+ca-b\frac{(\lambda^2-1-d)^2}{\lambda^2}-\tau^2\right]
+c(\tau^2-ab)=0,
\end{equation}
where 
\begin{eqnarray}
a&=&\lambda^2\left(1-\frac{v^2}{\lambda^2t^2}\right)-\frac{d}{2}%
\sin^22\theta,  \nonumber \\
b&=&1-\frac{v^2}{t^2}+\frac{d}{2}\sin^22\theta,  \nonumber \\
c&=&\frac{1}{\lambda^2}\left(1-\frac{v^2}{t^2}\right),  \nonumber \\
\tau&=&\frac{d}{4}\sin 4\theta.
\end{eqnarray}

``Weight factors'' in Eq. (\ref{fulsol}) have the following form: 
\begin{equation}  \label{pref}
\frac{C_{\alpha}^{(j)}}{C_{1}^{(j)}}=\left\{ 
\begin{array}{cc}
1, & \quad \mbox{for }\alpha=1, \\ 
-\frac{L_{21}}{L_{22}}\equiv p_2^{(j)}(\beta_j^2)= \frac{D/4\sin 4\theta}{%
-C_{44}\beta_j^2+C_{44}+D/2\sin^2 \theta-\rho v^2}, & \quad \mbox{for }%
\alpha=2, \\ 
-\frac{L_{31}}{L_{33}}\equiv \mp i\beta_j p_3^{(j)}(\beta_j^2)= \mp i\beta_j%
\frac{C_{11}-C_{44}-D}{C_{44}-C_{11}\beta_j^2-\rho v^2}, & \quad\mbox{for }%
\alpha =3,
\end{array}
\right.
\end{equation}
where the upper sign is for the upper medium, and for a convenience sake we
omitted primes in formulae for the lower medium. Then, for each term in the
sum (\ref{fulsol}) the condition (\ref{bc2}) on the stress continuity in
rotated frame can be written in the vector form: 
\begin{equation}  \label{sigv}
\left[ 
\begin{array}{c}
\sigma_{13}^j \\ 
\sigma_{23}^j \\ 
\sigma_{33}^j
\end{array}
\right] =i\left[ 
\begin{array}{ccc}
\pm i\beta_j C_{44} & 0 & C_{44} \\ 
0 & \pm i\beta_j C_{44} & 0 \\ 
C_{12} & 0 & \pm i\beta_j C_{11}
\end{array}
\right]\ \left[ 
\begin{array}{c}
1 \\ 
p_2^{(j)} \\ 
\mp i\beta_j p_3^{(j)}
\end{array}
\right].
\end{equation}
One can can see that the boundary conditions decouple for the sagittal plane 
$\left\{x_1x_3\right\}$ and the perpendicular direction $x_2$. With the help
of Eq. (\ref{sigv}) and minor simplifications we obtain six homogeneous
linear equations for the boundary conditions with the matrix ${\bf D}(v)$
equals to {\small 
\begin{equation}  \label{Dmatrix}
\left[ 
\begin{array}{cccccc}
1 & 1 & 1 & -1 & -1 & -1 \\ 
p_2^{(1)} & p_2^{(2)} & p_2^{(3)} & -p_2^{\prime(1)} & -p_2^{\prime(2)} & 
-p_2^{\prime(3)} \\ 
\beta_1 p_3^{(1)} & \beta_2 p_3^{(2)} & \beta_3 p_3^{(3)} & 
\beta_1^{\prime}p^{\prime}_3{^{(1)}} & \beta_2^{\prime}p_3^{\prime(2)} & 
\beta_3^{\prime}p_3^{\prime(3)} \\ 
C_{44}\beta_1p_2^{(1)} & C_{44}\beta_2p_2^{(2)} & C_{44}\beta_3p_2^{(3)} & 
C^{\prime}_{44}\beta^{\prime}_1p_2^{\prime(1)} & C^{\prime}_{44}\beta^{%
\prime}_2p_2^{\prime(2)} & C^{\prime}_{44}\beta^{\prime}_3p_2^{\prime(3)} \\ 
C_{44}\beta_1(1-p_3^{(1)}) & C_{44}\beta_2(1-p_3^{(2)}) & 
C_{44}\beta_3(1-p_3^{(3)}) & C^{\prime}_{44}\beta^{\prime}_1(1-p_3^{%
\prime(1)}) & C^{\prime}_{44}\beta^{\prime}_2(1-p_3^{\prime(2)}) & 
C^{\prime}_{44}\beta^{\prime}_3(1-p_3^{\prime(3)}) \\ 
C_{12}+C_{11}\beta_1^2p_3^{(1)} & C_{12}+C_{11}\beta_2^2p_3^{(2)} & 
C_{12}+C_{11}\beta_3^2p_3^{(3)} & -C^{\prime}_{12}-C^{\prime}_{11}\beta_1^{%
\prime 2}p_3^{\prime(1)} & -C^{\prime}_{12}-C^{\prime}_{11}\beta_2^{\prime
2}p_3^{\prime(2)} & -C^{\prime}_{12}-C^{\prime}_{11}\beta_3^{\prime
2}p_3^{\prime(3)}
\end{array}
\right].
\end{equation}
}

The condition for a nontrivial solution is that its determinant must vanish: 
\begin{equation}  \label{detvan}
D^{(\gamma)}(v;\beta_1,\ldots ,\beta^{\prime}_3)\equiv |D^{(\gamma)}_{ij}|=0.
\end{equation}
Since each $\beta$ is doubled value function of complex variable $v$, this
determinant has 64 Riemann sheets, which we denote by upper-script $\gamma$.

The sign convention for independent sheets is determined by real part of $%
\beta$. It is given as follows: 
\begin{equation}  \label{scon}
\mbox{sign}\left(\mbox{Re}(\beta_1),\mbox{Re}(\beta_2),\mbox{Re}(\beta_3), %
\mbox{Re}(\beta^{\prime}_1),\mbox{Re}(\beta^{\prime}_2),\mbox{Re}%
(\beta^{\prime}_3),\right).
\end{equation}
That is, Riemann sheet (+ + + - - -) corresponds to the case Re$(\beta_j)>0$
and Re$(\beta^{\prime}_j)<0$, where $j=1,2,3$.

The important property of the determinant is that its solutions are
invariant against the simultaneous change of sign of all six decay
constants. Indeed, the functions $p_2,\ p_3$ are defined in such a way that
they depend on $\beta^2$ only. Therefore, from the form of matrix (\ref
{Dmatrix}) it follows that simultaneous change of sign $\beta$ leaves the
secular equation unaltered. Thus, in order to find all the roots of Eq. (\ref
{detvan}) it is enough to investigate only 32 independent Riemann sheets. To
enumerate these sheets we, firstly, solve equation (\ref{efdetL}) and sort
obtained $\beta(v)$ for each medium in the following order: 
\[
|\mbox{Re}(\beta_1)|\le|\mbox{Re}(\beta_2)|\le|\mbox{Re}(\beta_3)|. 
\]
Secondly, we choose notation that the sign of the real part $\beta_1$ is
always positive. Then the upper Riemann sheet $\gamma=1$ corresponds to the
case (++++++) and the subsequent numbers for lower sheets are given in the
table III. 
\parbox{18cm}{
\begin{table}[h]\label{tab2}
\begin{tabular}{|c|cccccc|}
$\gamma$ & Re$\beta_1$ &Re$\beta_2$ &Re$\beta_3$ &Re$\beta'_1$  &
Re$\beta'_2$  &Re$\beta'_3$ \\
\hline
1&+&+&+&+&+&+\\
2&+&+&+&+&+&-\\
3&+&+&+&+&-&+\\
4&+&+&+&-&+&+\\
5&+&+&+&+&-&-\\
6&+&+&+&-&+&-\\
7&+&+&+&-&-&+\\
8&+&+&+&-&-&-\\
9&+&+&-&+&+&+\\
$\cdots$&$\cdots$&$\cdots$&$\cdots$&$\cdots$&$\cdots$&$\cdots$ \\
17&+&-&+&+&+&+\\
$\cdots$&$\cdots$&$\cdots$&$\cdots$&$\cdots$&$\cdots$&$\cdots$ \\
25&+&-&-&+&+&+\\
$\cdots$&$\cdots$&$\cdots$&$\cdots$&$\cdots$&$\cdots$&$\cdots$ \\
32&+&-&-&-&-&-\\
\end{tabular}
\end{table}
}

Another feature of Eq. (\ref{detvan}) is the following. If $v=v_R+iv_I$ is
the solution of the dispersion relation $D^{(\gamma)}(v)\equiv
D_R^{(\gamma)}+iD^{(\gamma)}_I=0$, then $v=v_R-iv_I$ will be also a solution
of Eq. (\ref{detvan}). Here we explicitly separated real and imaginary part
of complex function. It could be understood by the following consideration.
Let $\beta_l=\beta_{R_{l}}+i\beta_{I_{l}} (l=1,\ldots,6)$, obtained from
Eqs. (\ref{detL}) for $v=v_R+iv_I$, then the solutions of Eqs. (\ref{detL})
for $v=v_R-iv_I$ are $\beta_l=\beta_{R_l}-i\beta_{I_l}$. Moreover, the
simultaneous change $v\longrightarrow v_R-iv_I$, $\beta_l\longrightarrow
\beta_{R_l}-i\beta_{I_l}$ in Eq. (\ref{detvan}) brings up the change of sign
at the imaginary part of the determinant only: 
\[
D^{(\gamma)}(v_R-iv_I)=D^{(\gamma)}_R-iD^{(\gamma)}_I. 
\]
Thus $v=v_R-iv_I$ is also the solution.

If we suggest that frequency of the interface acoustic wave is real, then $%
k=\omega/v$ will be complex. We will choose solutions with $v=v_R-iv_I$
which correspond to the attenuated waves (for $\omega>0$) along direction of
its propagation on the interface.

The dispersion relation simplifies considerably for the propagation on the
interface (001) along the directions of high symmetry ([100] and [110]) and
in the isotropic case. For all this cases, matrix elements $%
L_{12}=L_{21}\equiv 0$ [because of $\sin 4\theta =0$ or $D=0$, see Eq. (\ref
{matrix})]. Therefore, the systems (\ref{homset}) for each media breaks up
into the pair: 
\begin{equation}  \label{pair}
\left[ 
\begin{array}{cc}
L_{11} & L_{13} \\ 
L_{13} & L_{33}
\end{array}
\right] \left[ 
\begin{array}{c}
A_1 \\ 
A_3
\end{array}
\right] =0
\end{equation}
and 
\begin{equation}
L_{22}A_2=0.
\end{equation}
The only nontrivial solution of the latter corresponds to a bulk transverse
acoustic wave propagating parallel to the interface of the elastic media 
\cite{Kress}, and therefore it is discarded. Thus, in these cases the
interface waves polarized in the sagittal plane and do not have $u_2$
component.

For the interface waves polarized in the sagittal  plane the solvability
condition for (\ref{pair}) is the biquadratic equation 
\begin{equation}  \label{biquad}
\lambda^2\beta^4-\beta^2\left(\gamma_1^2+\lambda^4 \gamma_2^2-
(\lambda^2-1-d)^2\right)+\lambda^2\gamma_1^2 \gamma_2^2=0,
\end{equation}
where we introduced notation 
\begin{eqnarray}
\lambda^2 &=& C_{11}/C_{44},  \label{n1} \\
d &=& D/C_{44},  \label{n2} \\
\gamma_1^2&=& 1-v^2/t^2,  \label{n3} \\
t^2 &=& C_{44}/\rho,  \label{n4}
\end{eqnarray}
and 
\begin{equation}  \label{g2}
\gamma_2^2=\left\{ 
\begin{array}{cc}
1-v^2/(\lambda t)^2 & \quad\mbox{for }\theta=0, \\ 
1-d/2\lambda^2-v^2/(\lambda t)^2 & \quad\mbox{for }\theta=\pi/4.
\end{array}
\right.
\end{equation}
In the isotropic case ($d=0$) the Eq. (\ref{biquad}) reduces to the 
\begin{equation}  \label{isotr}
\beta^4-\beta^2\left(\gamma_1^2+\gamma_2^2\right)-\gamma_1^2\gamma_2^2=0
\end{equation}
with obvious solutions 
\begin{eqnarray}  \label{betis}
\beta_1&=&\pm\sqrt{1-v^2/t^2}  \nonumber \\
\beta_2&=&\pm\sqrt{1-v^2/(\lambda t)^2}.
\end{eqnarray}
For all these cases the general solution for the interface wave are linear
combinations of two partial waves for both media: 
\begin{eqnarray}  \label{gsis}
u_1({\bf r},t)&=&\left[K_1 e^{-k\beta_1 x_3}+K_2 e^{-k\beta_2 x_3}\right]
e^{ik(x_1-vt)},  \nonumber \\
u_3({\bf r},t)&=&\left[K_1p_3^{(1)} e^{-k\beta_1 x_3}+K_2p_3^{(2)}
e^{-k\beta_2 x_3}\right]e^{ik(x_1-vt)},  \nonumber \\
u^{\prime}_1({\bf r},t)&=&\left[K^{\prime}_1 e^{k\beta^{\prime}_1
x_3}+K^{\prime}_2 e^{k\beta^{\prime}_2 x_3}\right] e^{ik(x_1-vt)},  \nonumber
\\
u^{\prime}_3({\bf r},t)&=&\left[K^{\prime}_1p_3^{\prime(1)}
e^{k\beta^{\prime}_1 x_3}+K^{\prime}_2 p_3^{\prime(2)} e^{k\beta^{\prime}_2
x_3}\right]e^{ik(x_1-vt)}.
\end{eqnarray}
After substitution solution (\ref{gsis}) into the boundary conditions (\ref
{bc1},\ref{bc2}) we obtain the set of four homogeneous linear equations with
solvability condition: 
\begin{equation}  \label{detDis}
\left| 
\begin{array}{cccc}
1 & 1 & -1 & -1 \\ 
\beta_1 p_3^{(1)} & \beta_2 p_3^{(2)} & \beta_1^{\prime}p_3^{\prime(1)} & 
\beta_2^{\prime}p_3^{\prime(2)} \\ 
C_{44}\beta_1(1-p_3^{(1)}) & C_{44}\beta_2(1-p_3^{(2)}) & 
C^{\prime}_{44}\beta^{\prime}_1(1-p_3^{\prime(1)}) & C^{\prime}_{44}\beta^{%
\prime}_2(1-p_3^{\prime(2)}) \\ 
C_{12}+C_{11}\beta_1^2p_3^{(1)} & C_{12}+C_{11}\beta_2^2p_3^{(2)} & 
-C^{\prime}_{12}-C^{\prime}_{11}\beta_1^{\prime 2}p_3^{\prime(1)} & 
-C^{\prime}_{12}-C^{\prime}_{11}\beta_2^{\prime 2}p_3^{\prime(2)}
\end{array}
\right|=0.
\end{equation}
For isotropic case $p_3^{(1)}=-1/\beta_1^2$ and $p_3^{(2)}=-1$. Then the
resulting dispersion relation reduces\cite{ston1} to 
\begin{equation}  \label{stdr1}
v^4\left[(\rho-\rho^{\prime})^2-(\rho\beta^{\prime}_2+\rho^{\prime}\beta_2)
(\rho\beta^{\prime}_1+\rho^{\prime}\beta_1)\right]+ 4Fv^2\left[%
\rho\beta^{\prime}_1\beta^{\prime}_2
-\rho^{\prime}\beta_1\beta_2-\rho+\rho^{\prime}\right] +4F^2(1-\beta_1%
\beta_2) (1-\beta^{\prime}_1\beta^{\prime}_2)=0,
\end{equation}
where 
\begin{equation}  \label{F}
F=\rho t^2-\rho^{\prime}t^{\prime 2}.
\end{equation}
There are {\em sixteen} independent roots on eight different Riemann sheets
for the given equation There is always non-attenuated solution for $%
t=t^{\prime}$, $\lambda=\lambda^{\prime}$ and $\rho\neq \rho^{\prime}$.

\section{GFSM method}\label{ApA}

In addition to our method described above it is also possible to obtain the
dispersion relation for the interface wave using Surface Green Function
Matching (SGFM) analysis (see for details Refs. \cite{GM,VGM}). 
We used this technique as a complementary independent method in our numerical
calculations to compare the results.
The main advantage of the method is that it gives the determinant of the 
matrix with dimensions 3x3, rather than 6x6. However, the calculation of 
the those matrix elements is more cumbersome. Another important development 
of this method is construction of the Green function of whole system, 
which also enables to calculate interface contribution to the important 
physical properties of the system such as the density of phonon modes, 
specific heat, and the atom mean square displacements.

We consider the (001) interface of cubic crystals and arbitrary propagation
directions. The idea of the method is to construct the Green function {\bf G}%
$^{{\bf s}}$ of the composite two-media system, such that it incorporates
the boundary conditions at the interface. In fact, to obtain the dispersion
relation it suffices to know the surface projection ${\bf g^s}$ of {\bf G}$^{%
{\bf s}}$ on the interface plane. Since the boundary conditions (\ref{bc1},%
\ref{bc2}) involve first derivatives, the surface projection of the bulk
Green functions ${\cal G}_{ij}$ and their normal derivatives will be
involved. These are defined by 
\begin{eqnarray}  \label{spgf}
G_{ij}&=&\lim_{x_3^{\prime}\rightarrow 0} \left[\langle x_3|{\cal G}%
_{ij}|x_3^{\prime}\rangle\right]_{x_3=0} =\lim_{\eta\rightarrow 0}\frac{1}{%
2\pi}\int_{-\infty}^{\infty}{\cal G}_{ij}(q) \exp(iq\eta)\,dq,  \nonumber \\
G_{ij}^{(\pm)}&=&\lim_{x_3^{\prime}\rightarrow \pm 0}\left[\frac{%
\partial\langle x_3|{\cal G}_{ij}|x_3^{\prime}\rangle}{\partial x_3} \right]%
_{x_3=0}= \lim_{\eta\rightarrow 0}\frac{1}{2\pi}\int_{-\infty}^{\infty}{\cal %
G}_{ij}(q) iq \exp(\mp iq\eta)\,dq.  \nonumber
\end{eqnarray}
In our case ${\bf {\cal G}}(q)=\left[{\bf L}\right]^{-1}$, where matrix {\bf %
L} is given by Eq. (\ref{matrix}) with substitution $i\beta\rightarrow q$.
The calculation is tedious but straightforward\cite{vel_unp}. 
\begin{eqnarray}
G_{11} & = &\sum_{j=1}^{3}\frac{ (\gamma_1^2-\beta_j^2\lambda^2)(\gamma_1^2+%
\frac{d}{2}\sin^2 2\theta -\beta_j^2)} {2\lambda^{2}\beta_{j}(%
\beta_{j+1}^{2}-\beta_{j}^{2}) (\beta_{j+2}^{2}-\beta_{j}^{2})C_{44}} , 
\nonumber \\
G_{22} & =&\sum_{j=1}^{3}\frac{ (\gamma_1^2-\lambda^2\beta_j^2)(\lambda^2-%
\frac{v^2}{t^2}-\frac{d}{2}\sin^2 2\theta-\beta_j^2)
+\beta_j^2(\lambda^2-1-d)^2} {2\lambda^{2}\beta_{j}(\beta_{j+1}^{2}-%
\beta_{j}^{2}) (\beta_{j+2}^{2}-\beta_{j}^{2})C_{44}} ,  \nonumber \\
G_{33} & = &\sum_{j=1}^{3}\frac{ (\gamma_1^2-\beta_j^2)(\lambda^2-\frac{v^2}{%
t^2}-\beta_j^2)+\frac{d}{2}\sin^2 2\theta (-\frac{d}{2}+\lambda^2-1)} {%
2\lambda^{2}\beta_{j}(\beta_{j+1}^{2}-\beta_{j}^{2})(\beta_{j+2}^{2}-%
\beta_{j}^{2})C_{44}},  \nonumber \\
G_{12} & =&{G}_{21}=\frac{d\tau}{2\lambda^{2}}\sum_{j=1}^{3}\frac{%
\lambda^{2}\beta_{j}^{2}- \gamma_1^2} {\beta_{j}(\beta_{j+1}^{2}-%
\beta_{j}^{2})(\beta_{j+2}^{2}-\beta _{j}^{2})C_{44}},  \nonumber \\
G_{13} & =&G_{31}=G_{23}=G_{32}=0;  \nonumber \\
G_{11}^{(\pm)} & = & G_{22}^{(\pm)}= \pm\frac{1}{2C_{44}},  \nonumber \\
G_{33}^{(\pm)} & = &\pm\frac{1}{2C_{11}},  \nonumber \\
G_{12}^{(\pm)} & = &G_{21}^{(\pm)}=0,  \nonumber \\
G_{13}^{(\pm)} & = &G_{31}^{(\pm)}= -\frac{i(\lambda^{2}-1-d)}{%
2\lambda^{2}C_{44}}\sum_{j=1}^{3}\frac{\left[ \beta_{j}^{2}-(1+\frac{d}{2}%
\sin^{2}2\theta)+\frac{v^{2}}{t^{2}}\right]\beta_{j}}{(\beta_{j+1}^{2}-%
\beta_{j}^{2})(\beta _{j+2}^{2}-\beta_{j}^{2})},  \nonumber \\
G_{23}^{(\pm)} & =&G_{32}^{(\pm)}= -\frac{id\tau(\lambda^{2}-1-d)}{%
2\lambda^{2}C_{44}}\sum_{j=1}^{3}\frac{\beta_{j}}{(\beta_{j+1}^{2}-%
\beta_{j}^{2})(\beta_{j+2}^{2}-\beta_{j}^{2})}.
\end{eqnarray}
Here $\theta$ is the angle formed by the direction of propagation and the $%
x_{1}$ axis, $\tau=-\frac{\sin4\theta}{4}$ and notations for $%
\lambda,d,\gamma_1, t$ are given by Eqs. (\ref{n1}-\ref{n4}). The $\beta_{1}$%
, $\beta_{2}$ and $\beta_{3}$ are the solutions of Eq. (\ref{detL}). It is
proved \cite{GM} that ${\bf g^{S}}$ is given by 
\begin{equation}  \label{gs}
[{\bf g^{S}}]^{-1}={\bf A}^{(+)}{\bf G}^{-1}-{\bf A}^{\prime(-)}{\bf G}%
^{\prime -1},
\end{equation}
where the inverse is defined in the two-dimensional space of the interface.
The matrix ${\bf A}$ is a suitably defined linear operator. Its appearance
in Eq. (\ref{gs}) reflects the boundary conditions. 
\begin{equation}
{\bf A}^{(\pm)}=\left[ 
\begin{array}{ccc}
C_{44}{G}_{11}^{(\pm)} & 0 & C_{44}(G_{13}^{(\pm)}+i G_{33}) \\ 
0 & C_{44}{G}_{22}^{(\pm)} & C_{44}{G}_{23}^{(\pm)} \\ 
C_{12}i {G}_{11}+C_{11}{G}_{31}^{(\pm)} & C_{12}i {G}_{12}+C_{11}{G}%
_{32}^{(\pm)} & C_{11}{G}_{33}^{(\pm)}
\end{array}
\right]
\end{equation}
The secular equation, which gives the interface mode dispersion relation is
shown to be 
\begin{equation}  \label{drgfsm}
\mbox{det}\,\left[{\bf g^s}\right]^{-1}=0
\end{equation}

For high-symmetry directions the further simplifications are possible. In
this case we separate the transverse bulk mode again and the resulting
matrix $[{\bf g^S}]^{-1}$ for interface waves is two-dimensional with the
following matrix elements: 
\begin{eqnarray}
\left[g^s\right]^{-1}_{11} & = & \frac{C_{44}\lambda^{2}\beta_{1}\beta_{2}(%
\beta_{1}^{2}-\beta_{2}^{2})} {\beta_{2}(\lambda^{2}\beta_{1}^{2}-%
\gamma_1^2) -\beta_{1}(\lambda^{2}\beta_{2}^{2}-\gamma_1^2)} +\frac{%
C_{44}^{^{\prime}}\lambda^{\prime2}\beta_{1}^{^{\prime}}\beta_{2}^{^{%
\prime}} (\beta_{1}^{\prime2}-\beta_{2}^{^{\prime}2})} {\beta_{2}^{^{%
\prime}}(\lambda^{\prime2}\beta_{1}^{\prime2}-\gamma_1^{\prime2})
-\beta_{1}^{^{\prime}}(\lambda^{\prime2}\beta_{2}^{\prime2}-\gamma_1^{%
\prime2})}  \nonumber \\
\left[g^s\right]^{-1}_{22} & = & \frac{C_{44}\lambda^{2}\beta_{1}\beta_{2}(%
\beta_{1}^{2}-\beta_{2}^{2})} {\beta_{2}(\beta_{1}^{2}-\lambda^{2}%
\gamma_2^2) -\beta_{1}(\beta_{2}^{2}-\lambda^{2}\gamma_2^2)} + \frac{%
C_{44}^{^{\prime}}\lambda^{\prime2}\beta_{1}^{^{\prime}}\beta_{2}^{^{%
\prime}} (\beta_{1}^{\prime2}-\beta_{2}^{^{\prime}2})} {\beta_{2}^{^{%
\prime}}(\beta_{1}^{\prime2}-\lambda^{\prime2}\gamma_2^{\prime2})
-\beta_{1}^{^{\prime}}(\beta_{2}^{\prime2}-\lambda^{\prime2}\gamma_2^{%
\prime2})}  \nonumber \\
\left[g^s\right]^{-1}_{12} & = & i \left[\frac{C_{44}(\lambda^{2}-1-d)%
\beta_{1}\beta_{2}(\beta_{2}-\beta_{1})} {\beta_{2}(\beta_{1}^{2}-%
\lambda^{2}\gamma_2^2) -\beta_{1}(\beta_{2}^{2}-\lambda^{2}\gamma_2^2)}%
+C_{44} -\frac{C_{44}^{\prime}(\lambda^{\prime2}-1-d^{^{\prime}})
\beta_{1}^{^{\prime}}\beta_{2}^{^{\prime}}(\beta_{2}^{\prime}-\beta_{1}^{^{%
\prime}})} {\beta_{2}^{^{\prime}}(\beta_{1}^{\prime2}-\lambda^{\prime2}%
\gamma_2^{\prime2})
-\beta_{1}^{^{\prime}}(\beta_{2}^{\prime2}-\lambda^{\prime2}\gamma_2^{%
\prime2})} -C_{44}^{\prime}\right]  \nonumber \\
\left[g^s\right]^{-1}_{21} & = & i\left[\frac{C_{11}(\lambda^{2}-1-d)%
\beta_{1}\beta_{2}(\beta_{2}-\beta_{1})} {\beta_{2}(\lambda^{2}%
\beta_{1}^{2}-\gamma_1^2) -\beta_{1}(\lambda^{2}\beta_{2}^{2}-\gamma_1^2)}%
+C_{12} -\frac{C_{11}^{^{\prime}}(\lambda^{\prime2}-1-d^{^{\prime}})
\beta_{1}^{^{\prime}}\beta_{2}^{^{\prime}}
(\beta_{2}^{\prime}-\beta_{1}^{^{\prime}})} {\beta_{2}^{^{\prime}}(\lambda^{%
\prime2}\beta_{1}^{\prime2}-\gamma_1^{\prime2})
-\beta_{1}^{^{\prime}}(\lambda^{\prime2}\beta_{2}^{\prime2}-\gamma_1^{%
\prime2})} -C_{12}^{^{\prime}}\right]  \nonumber
\end{eqnarray}
$\beta_{1}$ and $\beta_{2}$ are the solutions of Eq. (\ref{biquad}) with $%
\gamma_2$ defined by Eq. (\ref{g2}). The dispersion relation for the
interface waves can be obtained from the zeros of det$[{bf g^s}]^{-1}$.

The former expression for the determinant can be reduced in a very easy way
to the relationships for the isotropic case. When $D=C_{11}-C_{12}-2C_{44}=0$
the matrix elements are given by:

\begin{eqnarray}  \label{gfsmis2}
\left[g^s\right]^{-1}_{11} &= &\frac{v^2\left[ \rho\beta_2(1-\beta^{%
\prime}_1\beta^{\prime}_2)+\rho^{\prime}\beta^{\prime}_2(1-\beta_1\beta_2)%
\right]} {(1-\beta_1\beta_2)(1-\beta^{\prime}_1\beta^{\prime}_2)},  \nonumber
\\
\left[g^s\right]^{-1}_{22} &= &\frac{v^2\left[ \rho\beta_1(1-\beta^{%
\prime}_1\beta^{\prime}_2)+\rho^{\prime}\beta^{\prime}_1(1-\beta_1\beta_2)%
\right]} {(1-\beta_1\beta_2)(1-\beta^{\prime}_1\beta^{\prime}_2)},  \nonumber
\\
\left[g^s\right]^{-1}_{12} &=& -\left[g^s\right]^{-1}_{21} =i\left[%
2F-v^2\left(\frac{\rho}{1-\beta_1\beta_2}- \frac{\rho^{\prime}}{%
1-\beta^{\prime}_1\beta^{\prime}_2}\right)\right].
\end{eqnarray}

\end{document}